\documentclass[journal]{vgtc}                     

\onlineid{1121}

\vgtccategory{Research}

\vgtcpapertype{application/design study}

\title{PhenoFlow: A Human-LLM Driven Visual Analytics System for Exploring Large and Complex Stroke Datasets}

\author{
    \authororcid{Jaeyoung Kim}{0009-0006-1868-7148}, \authororcid{Sihyeon Lee}{0009-0006-7202-234X}, \authororcid{Hyeon Jeon}{0000-0002-9659-2922}, Keon-Joo Lee, Hee-Joon Bae, \authororcid{Bohyoung Kim}{0000-0002-2183-5651}, and \authororcid{Jinwook Seo}{0000-0002-7734-822X} 
}

\authorfooter{
  \vspace{-1mm}
  \item
  	Jaeyoung Kim, Sihyeon Lee, Hyeon Jeon, and Jinwook Seo are with Seoul National University.
  	E-mail: \{jykim, sihyeon, hj\}@hcil.snu.ac.kr, jseo@snu.ac.kr 
    \item 
    Keon-Joo Lee is with the Department of Neurology, Korea University Guro Hospital. E-mail: gooday19@gmail.com
    \item 
    Hee-Joon Bae is with the Seoul National University College of Medicine and Seoul National University Bundang Hospital. E-mail: braindoc@snu.ac.kr
    \item 
     Bohyoung Kim (corresponding author) is with Hankuk University of Foreign Studies. E-mail: bkim@hufs.ac.kr
}

\abstract{
Acute stroke demands prompt diagnosis and treatment to achieve optimal patient outcomes. However, the intricate and irregular nature of clinical data associated with acute stroke, particularly blood pressure (BP) measurements, presents substantial obstacles to effective visual analytics and decision-making. Through a year-long collaboration with experienced neurologists, we developed PhenoFlow, a visual analytics system that leverages the collaboration between human and Large Language Models (LLMs) to analyze the extensive and complex data of acute ischemic stroke patients. PhenoFlow pioneers an innovative workflow, where the LLM serves as a data wrangler while neurologists explore and supervise the output using visualizations and natural language interactions. This approach enables neurologists to focus more on decision-making with reduced cognitive load. To protect sensitive patient information, PhenoFlow only utilizes metadata to make inferences and synthesize executable codes, without accessing raw patient data. This ensures that the results are both reproducible and interpretable while maintaining patient privacy. The system incorporates a slice-and-wrap design that employs temporal folding to create an overlaid circular visualization. Combined with a linear bar graph, this design aids in exploring meaningful patterns within irregularly measured BP data. Through case studies, PhenoFlow has demonstrated its capability to support iterative analysis of extensive clinical datasets, reducing cognitive load and enabling neurologists to make well-informed decisions. Grounded in long-term collaboration with domain experts, our research demonstrates the potential of utilizing LLMs to tackle current challenges in data-driven clinical decision-making for acute ischemic stroke patients.
\vspace{-0.5mm}
}

\keywords{Stroke, Irregularly spaced time-series data, Multi-dimensional data, Cohort analysis, Large language models
\vspace{-3mm}}

\teaser{
  \centering
  \vspace{-2mm}
  \includegraphics[width=\linewidth]{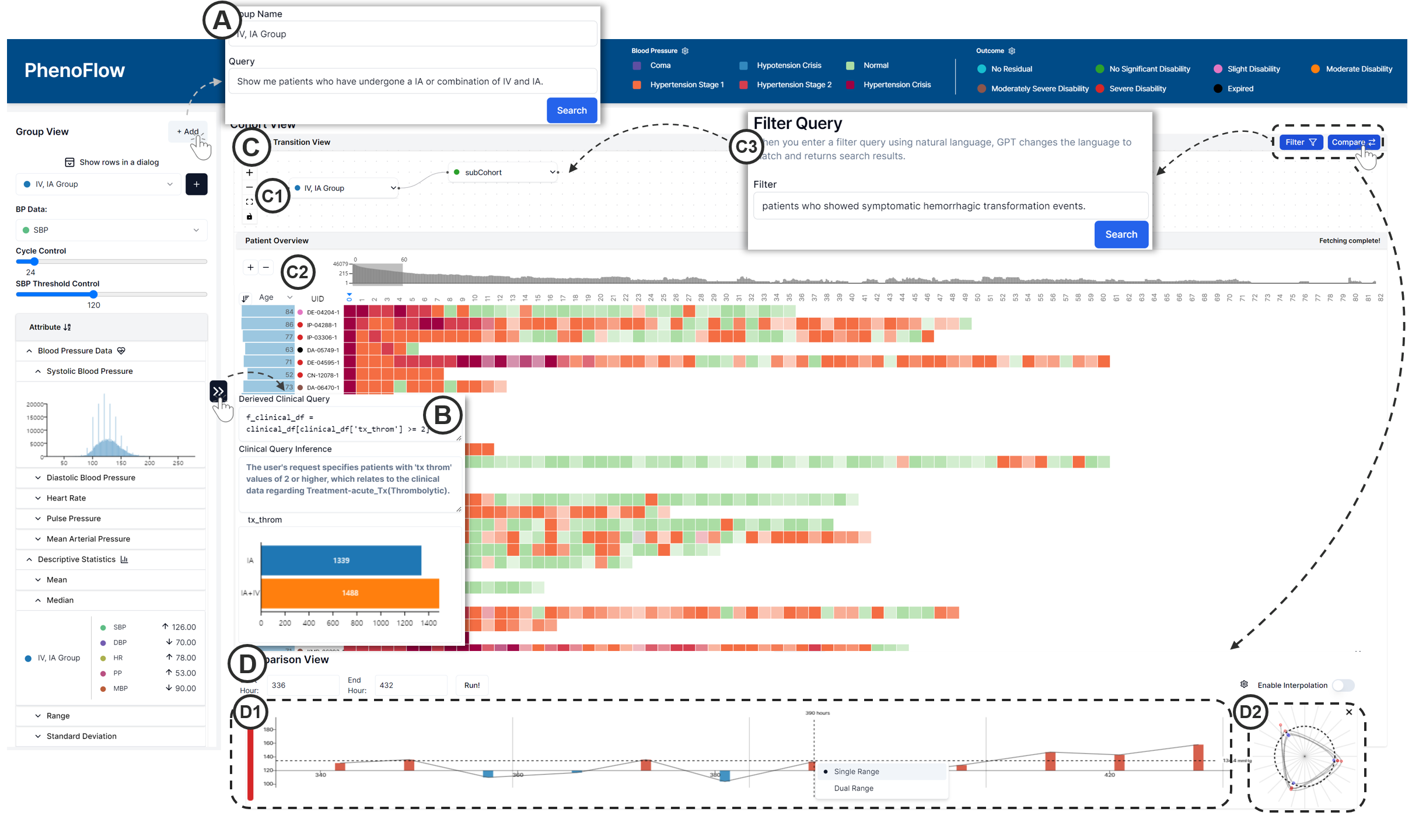}
  \vspace{-9mm}
  \caption{PhenoFlow empowers neurologists to explore large and complex medical datasets for acute ischemic stroke patients, with reduced cognitive load. (A) The cohort construction component allows neurologists to define a target cohort using natural language queries. (B) The Visual Inspection View provides explanations in plain language and small multiples of relevant fields to support understanding and debugging the behavior of the LLM data wrangler. (C) The Cohort View summarizes (C1) the cohort relationships in a node-link diagram and (C2) each patient's blood pressure (BP) trajectories as a matrix visualization. (C3) The natural language filtering supports iterative exploration of the defined cohorts. The (D1) linear bar charts and (D2) slice-and-wrap visualization present individual BP trajectories as time-series, revealing distinct patterns like triangular shapes in the irregularly measured BP data.
  }
  \vspace{-2mm}
}

\graphicspath{{figs/}{figures/}{pictures/}{images/}{./}} 

\usepackage{tabu}                      
\usepackage{booktabs}                  
\usepackage{lipsum}                    
\usepackage{mwe}                       
\usepackage{tabularx}
\usepackage{booktabs}
\usepackage{multirow}

\usepackage{mathptmx}                  

\begin{document}


\maketitle
\section{Introduction}
Acute stroke, characterized by the sudden obstruction of blood vessels in the brain, demands prompt decision-making to improve treatment efficacy and reduce long term disability. The widespread adoption of Electronic Medical Records (EMRs) has significantly expanded the volume of patient data available for stroke research \cite{murdoch2013inevitable}. However, this rapid data growth has also amplified the presence of legacy data, human errors and inconsistencies.  Data pertaining to stroke cases often feature irregular time intervals, inconsistent terminologies, and a lack of standardized structure across organizations. Consequently, these factors contribute to the creation of large and complex datasets, posing significant challenges for the effective and efficient analysis of data.

While a growing body of visual analytics research focuses on addressing these issues, there remains a notable gap in visual analytics tools specifically designed for time-sensitive diseases such as acute ischemic stroke. Previous efforts, like Stroscope \cite{cho2014stroscope} and TimeSpan \cite{loorak2015timespan}, made remarkable strides in handling data irregularity and supporting data analysis. However, as data complexity and volume have increased, modern clinical datasets now encompass data from tens to hundreds of thousands of unique patients \cite{bae2022david, johnson2016mimic, johnson2020mimic}. This scale highlights the need for more advanced visual analytics systems and novel analysis workflows tailored to the challenges of large, complex acute ischemic stroke data. To address these challenges, we collaborated with domain experts to identify bottlenecks in the existing analysis workflow. We found that neurologists often struggle with the cognitive load imposed by large datasets \cite{caban2015visual} and visual clutter, even in familiar visual representations (e.g., bars and lines). The prevailing analysis workflows, which rely heavily on neurologists' cognitive abilities, are inadequate for exploring such large, complex datasets, leading to prolonged analysis times and potentially overlooking critical patterns and insights.

Based on this understanding, we established three key domain goals: (1) facilitating cohort construction and exploration, (2) supporting the discovery of meaningful temporal patterns, and (3) providing clinical evidence alongside derived outcomes. Additionally, we identified the crucial importance of preserving patient privacy while working with sensitive medical data. These goals, along with seven key design requirements (see \hyperref[sec:design_requirements]{Sec. 4.5}), informed the design and implementation of PhenoFlow, a visual analytics system driven by human-Large Language Model (LLM) collaboration to explore large, complex acute ischemic stroke data. PhenoFlow introduces a novel analysis workflow, employing LLMs for cohort construction and facilitating an iterative exploration process. This workflow empowers neurologists to effectively explore and analyze vast and complex medical datasets with reduced cognitive load. To address the privacy concerns, PhenoFlow employs a novel approach that utilizes metadata to generate inferences, synthesize executable code for cohort construction, and create visualizations without directly accessing raw patient data. To mitigate potential errors generated by the LLM, PhenoFlow incorporates a visual inspection view that allows users to debug the intermediate results and validate the final LLM-derived output. This visual feedback mechanism, which includes visualizations of data distributions, helps to ensure the accuracy and reliability of the results (see \hyperref[sec:LLMs]{Sec. 5.1}).

We also introduce the slice-and-wrap visualization technique, designed specifically to address the challenges posed by unevenly spaced time series data. This technique leverages temporal folding \cite{du2016coping} to split temporal points into segments based on predefined intervals or biological cycles. Each segment is then 'wrapped' in a circular visualization, which is superimposed to facilitate the exploration of recurring patterns. We opt for a circular layout for three reasons: (1) it displays lengthy sequences in a space-efficient way \cite{nusrat2019tasks}, (2) it aligns with the periodic characteristics inherent in blood pressure (i.e., biological clock), and (3) it avoids excessive visual clutter. Moreover, overlaying of information on tracks enables efficient comparison and summarization \cite{lyi2020comparative}. However, the identification of not only recurring patterns but also abnormal patterns is important to support decision-making in acute ischemic stroke scenarios. To intuitively reveal less frequent yet abnormal patterns, we juxtapose linear layout bar charts with interactive baselines. This combination proves adept at revealing both recurring and abnormal patterns in irregular data (see \hyperref[sec:circular_vis]{Sec. 5.2}).

To validate the effectiveness of PhenoFlow, we conducted a series of case studies. We engaged neurologists to use the system and analyze real-world patient data from the CRCS-K dataset \cite{bae2022david}. The results demonstrate that our visual analytics approach enhances physicians' ability to derive meaningful insights and increases their capacity for navigating complex, large datasets with minimal cognitive load.

While the case studies demonstrated promising results, we acknowledge limitations and areas for future research. One key issue raised by neurologists during our case studies is the challenge of validating findings generated through visual analytics tools. To address this, we propose developing techniques that enhance the visual interpretation skills of domain experts and enable them to better understand and trust the insights derived from visualizations. Another limitation relates to the fragility of LLMs as data wranglers. While LLMs have shown impressive data wrangling capabilities, ensuring their consistent and reliable behavior remains a challenge. We discuss potential approaches to mitigate this issue. Based on these insights, we propose several future directions for developing visual analytics tools that incorporate LLMs for exploring large, complex medical datasets (see \hyperref[sec:future_direction]{Sec. 7.1}).

In summary, our main contributions are:
\begin{itemize}
\item Three domain goals and seven design requirements identified through close collaboration with domain experts.
\item PhenoFlow, a visual analytics tool driven by human-LLM collaboration for exploring large, complex medical data. It integrates several LLM-based techniques with novel visualization approaches.
\item Case studies that demonstrate PhenoFlow's strengths, limitations, ability to discover meaningful patterns and trends in acute ischemic stroke data.
\end{itemize}
\vspace{-2mm}
\section{Related Work}
\vspace{-0.5mm}
This section provides an overview of the prior research that has influenced the design of PhenoFlow. We first discuss previous studies on visualizing time-oriented data, including stroke patient data, highlighting their contributions and limitations in handling large, complex datasets (\hyperref[sec:Time]{Sec. 2.1}). We then delve into the growing adoption of LLMs in the clinical domain, examining their applications, challenges, and potential for enhancing medical research workflows (\hyperref[sec:LLM-Related]{Sec. 2.2}). 

\vspace{-1mm}
\subsection{Time-Oriented Data and Stroke Visualization} \label{sec:Time}
\vspace{-0.5mm}
Analyzing time-oriented data, encompassing both time-series data and temporal event sequences, is an important task in medical decision-making scenarios. However, time-oriented data in the clinical field often suffer from irregularity and uncertainty in both data and temporal dimensions \cite{aigner2007visualizing}. In particular, blood pressure (BP) data, which is primary data in stroke research, is measured at irregular time intervals and does not necessarily follow a regular, predictable schedule.

A straightforward and effective method for tackling such data is transforming time-series data to temporal event sequences (i.e., \textbf{temporal abstraction}) \cite{batal2009temporal, federico2014qualizon, lammarsch2013interactive, moskovitch2009medical}. By doing so, irregular time-series data can be converted to discrete observations collected over time and arranged in sequence \cite{guo2021survey}. These temporal event sequences can be visualized with respect to the time axis. Lifelines \cite{plaisant2003lifelines} employed this approach. However, when analyzing large datasets, this approach can incur substantial cognitive load. To reduce the complexity, LifeLines2 \cite{wang2009temporal} and IDMVis \cite{zhang2018idmvis} adopted \textbf{temporal folding}, which involves folding or splitting long data streams into daily, weekly, monthly, or yearly segments to find cyclic patterns \cite{du2016coping}. Similar to prior work, PhenoFlow employs both temporal abstraction and temporal folding approaches to construct each patient's BP sequence at a cohort level. We aggregate each patient's BP measurements within a 24-hour segment (i.e., circadian rhythm) to efficiently summarize their BP trajectories. However, these approaches come with a loss of information. Therefore, we also provide support for visualizing BP data as time-series data at an individual level to ensure comprehensive analysis.

{Recent research by Scheer et al.\cite{scheer2022visualization} provides a comprehensive overview of visualization techniques for time-oriented medical data. The authors identify strategies employed in existing tools, such as juxtaposition, superposition, and explicit encoding. For example, } to visually address irregular time-series data (i.e., BP data), Stroscope \cite{cho2014stroscope} proposed a ripple graph that maps uncertainty between two temporal measurements by varying color intensity. Another approach that deals with irregularity is using animation. TimeRider \cite{rind2011visually} reveals temporal aspects using an animated scatter plot. Inspired by these prior works, we developed two visualizations to handle irregular BP data. At the cohort level, we mapped the data density of each segment of BP sequences to opacity. At the individual level, we propose a novel slice-and-wrap design that superimposes multiple circular visualizations to support the visual exploration of recurring patterns. This visualization is juxtaposed with a linear bar graph, which facilitates the exploration of less frequent but abnormal patterns, offering a comprehensive view of the individual patient's BP data.

Another study that helps clinicians visually analyze stroke patient data is TimeSpan \cite{loorak2015timespan}. TimeSpan supports the exploration of both multi-dimensional and temporal attributes of acute ischemic stroke patients. Similar to TimeSpan \cite{loorak2015timespan} and Stroscope \cite{cho2014stroscope}, our solution's primary goal is to support the exploration of both multi-dimensional and temporal data of acute ischemic stroke patients. However, PhenoFlow goes beyond prior work by enabling the exploration of voluminous and complex clinical datasets (see \hyperref[sec:crcsk]{Sec. 3.1}). By leveraging a large language model as a data wranger, PhenoFlow empowers neurologists to efficiently navigate and derive insights from extensive, multifaceted stroke datasets, addressing the challenges posed by the increasing scale and complexity of modern clinical data.

\vspace{-1mm}
\subsection{LLMs for Clinical Research} \label{sec:LLM-Related}
\vspace{-0.5mm}
The adoption of natural language processing (NLP) techniques in the medical domain has significantly increased in recent years. Especially after the introduction of GPT-3 \cite{brown2020language} and GPT-4 \cite{achiam2023gpt}, several LLMs have emerged that can achieve expert-level performance in the medical domain \cite{singhal2023towards, singhal2023large}. Concurrent with this trend, we explored opportunities in the clinical data visualization domain to reduce the cognitive load on neurologists during the analysis process and enhance their efficiency with the human-LLM collaborative workflow. He et al. classified LLMs' fundamental tasks in the healthcare domain into six areas: Named Entity Recognition (NER), Relation Extraction (RE), Text Classification (TC), Semantic Textual Similarity (STS), Question Answering (QA), and Dialog \cite{he2023survey}. Among these areas, PhenoFlow utilizes LLMs for STS and QA, which are part of the data wrangling tasks.

STS evaluates the extent to which two phrases or sentences convey the same meaning. In the clinical domain, STS is often used to address challenges related to inconsistent terminologies and human errors in electronic health records (EHRs). The National NLP Clinical and BioCreative/Open Health NLP challenge \cite{rastegar2018biocreative} demonstrated that STS can help reduce mistakes and disorganization in medical datasets. Our dataset also includes human errors and inconsistent terminologies introduced during the data transcription and integration process. These obstacles not only hinder the interpretability of data but also generate substantial cognitive loads for neurologists during the data wrangling.

QA typically refers to a task that involves generating or retrieving answers for given questions. Traditional QA systems often encounter challenges within the medical domain, primarily stemming from the vast amount of domain-specific knowledge \cite{guo2022medical}. However, with the advent of powerful LLMs such as GPT-4 \cite{achiam2023gpt}, prompting and in-context learning-based QA systems have been developed. Hemidi et al. \cite{hamidi2023evaluation} evaluated the performance of ChatGPT, Google Bard, and Claude for patient QA tasks from EHRs. Similarly, several prior studies report the QA performance of GPT-3.5 and GPT-4 in various medical domains such as dementia, bariatric, and general surgery \cite{wang2023can, samaan2023assessing, oh2023chatgpt}. While LLMs show impressive results, there are mixed findings in the literature. Some studies suggest that LLMs outperform medical experts \cite{holmes2023evaluating}, while others indicate that LLMs do not significantly surpass experts \cite{duong2023analysis}. Moreover, modern LLMs such as GPT-4, Claude, and Bard maintain proprietary architectures, posing challenges for quantitative performance evaluation in many uncertain areas.

To overcome the hurdles in medical data analysis and evaluate the performance of LLMs in this domain, we adopted GPT-4 with few-shot prompting, multi-step reasoning, and self-reflection. We tested the performance of LLMs for multiple medical research tasks and then finalized the design of the workflow, which leverages the strengths of LLMs while mitigating their limitations through human supervision and visual feedback mechanisms. The proposed workflow aims to reduce the cognitive load on neurologists during the data analysis process and enhance the efficiency of the analysis. We will discuss our approach and results in detail in later section (see \hyperref[sec:LLMs]{Sec 5.1}).

Recent research has employed LLMs for data wrangling \cite{narayan2022can} and processing SQL queries \cite{trummer2022codexdb}. In the clinical domain, EHRAgent \cite{shi2024ehragent}, LeafAI \cite{dobbins2023leafai}, and quEHRy \cite{soni2023quehry} have generated queries with NLP models and LLM agents from structured medical datasets. However, the primary goal of prior work is to evaluate accuracy and increase Text2SQL performance. We focus on the design of the human-LLM collaborative workflow that ensures reproducibility and explainability. We will later discuss the roles of visualization, LLMs, and experts in this workflow.

Lastly, preserving patient privacy is crucial when applying LLMs to the clinical domain \cite{qiu2023large}. Despite the anonymization and de-identification of much medical data, prior studies have revealed vulnerabilities where attackers can extract information by using specific prompts \cite{carlini2021extracting, carlini2022quantifying}. To mitigate this risk, we propose a novel approach that utilizes LLMs to make inferences based on metadata. Subsequently, the system synthesizes code capable of constructing cohort, implementing filtering, and generating visual inspections, thereby safeguarding patient privacy while facilitating robust data analysis.
\vspace{-1.5mm}
\section{Background}
\vspace{-0.5mm}
This section provides an explanation on a conceptual overview of acute ischemic stroke and the related data that PhenoFlow aims to support. We introduce the CRCS-K dataset \cite{bae2022david} and describe its key characteristics that informed the design of PhenoFlow.

\vspace{-1mm}
\subsection{Collaborator and Dataset Description} \label{sec:crcsk}
\vspace{-0.5mm}
We utilized the \textbf{CRCS-K} dataset \cite{bae2022david}, a multicenter cohort dataset containing 324 clinical variables from about 100,000 acute ischemic stroke patients, collected from 17 university hospitals in South Korea since 2011. Multiple groups of neurologists have reviewed the data, ensuring its credibility and relevance for stroke research. The dataset is segmented into five distinct parts, encompassing demographic information, risk factors, treatment and examination details, follow-up data, and vital BP measurements. 

Given the dataset's extensive longitudinal span and scale, careful curation of data is necessary. Neurologists \textbf{E1} and \textbf{E2}, co-authors of this paper and experts in acute ischemic stroke from Seoul National University Bundang Hospital and Korea University Guro Hospital, respectively, have led the data curation process. With 35 and 24 years of experience in the field, respectively, they selected 60 clinical variables and 5 BP-related variables crucial for acute ischemic stroke research. This meticulous selection process supports PhenoFlow's development, ensuring it effectively meets the nuanced demands of acute ischemic stroke research.

\vspace{-1mm}
\subsection{Acute Ischemic Stroke}
\vspace{-0.5mm}
Acute ischemic stroke, which accounts for approximately 85\% of all stroke cases \cite{adams1993classification}, occurs when a blood clot or other obstruction blocks blood flow to the brain, leading to rapid cell death and neurological deficits \cite{albers2002transient}. In acute ischemic stroke, when BP surpasses 150-160 mmHg for mean arterial pressure, the autoregulation mechanism that normally maintains stable cerebral blood flow may become impaired. This impairment can lead to unstable cerebral blood flow, potentially worsening brain damage \cite{jauch2013guidelines}. Consequently, BP serves as a crucial indicator for assessing patients' conditions \cite{lawes2004blood, lindenstrom1995influence}. Moreover, clinical events such as treatment interventions (e.g., thrombolysis or thrombectomy) and stroke recurrence contribute to understanding the underlying reasons for fluctuations in patients' BP.

\vspace{-1mm}
\subsection{Data Characteristics}
\vspace{-0.5mm}
PhenoFlow incorporates three key types of data: event data, clinical data, and BP data.

\textbf{Event data} capture significant medical events within patients' records, such as Intravenous Thrombolysis (IVT), Intra-Arterial Thrombolysis (IAT), and stroke recurrence. These events are discrete data paired with time. They can be recorded not only at specific points but also as intervals, indicating events that happen over periods of time (e.g., IA surgery from start to end).

\textbf{Clinical data} include discrete clinical features, such as the Modified Rankin Score (mRS), risk factors, demographics, medications, and the Trial of Org 10172 in Acute Stroke Treatment (TOAST) classification. These data are all categorical and remain unchanged over time, representing static entries at the time of collection.

\textbf{BP (Blood pressure) data } record hemodynamic information from patients' onset to discharge. These measurements are typically taken at irregular time intervals, which can lead to complexity in analysis. To address this issue, neurologists often aggregate BP measurements using a fixed time interval, such as the circadian rhythm (24-hour cycle). Our collaborators specifically requested to follow the circadian rhythm when starting the analysis. As a result, PhenoFlow was designed to aggregate and visualize BP data in 24-hour increments by default.

Depending on the analytical approach, BP measurements can be treated as either time-series or temporal event sequences. When the fluctuation of BP values (i.e., patterns) is important, BP data are treated as time-series data. The most commonly studied patterns, such as repeated spikes, sustained high BP, and sharp drops \cite{mistry2019blood}, typically indicate a negative health transition in the patient.

Conversely, when understanding the general health transition pattern of multiple patients, BP data are treated as temporal event sequences. For example, most acute ischemic stroke patients are hospitalized with high BP ranges, typically greater than 160 mmHg for systolic blood pressure (SBP). Neurologists may want to construct a cohort that shares similar clinical conditions, and analyze BP trajectories of patients in the cohort from admission to discharge. This exploration can be related to certain clinical questions, such as \textit{"Given patients who share similar clinical conditions, should clinicians aggressively lower BP to target levels using medication, or would a conservative approach of waiting for natural BP normalization be more advisable?"}

PhenoFlow's design takes into account these characteristics of BP data. At a cohort level, PhenoFlow treats BP measurements as temporal event sequences to support a more concise exploration. At an individual level, PhenoFlow addresses BP measurements as time-series data to support pattern analysis and comparison.
\vspace{-1mm}
\section{Problem Definition and Design}
\vspace{-1mm}
In this section, we describe the design process and present the domain goals and design requirements. We characterized the problem at a domain level and connected it to specific visual analysis tasks.

\vspace{-1mm}
\subsection{Design Process}
\vspace{-1mm}
Our goal is to develop a visual analytics system to support neurologists. To understand their expectations and needs, we conducted a design process based on the framework established by Sedlmair \textit{et al.}\cite{sedlmair2012design}. The process was divided into two primary phases.\\
\textbf{(Phase I) Initial exploration and domain analysis.}
In the first phase, our aim was to understand the current workflow of the experts and identify their specific needs. Over a span of six months, we conducted weekly meetings with two neurologists. We observed their existing workflows and comprehensively reviewed domain literatures pertaining to acute ischemic stroke. Based on the insights collected from this process, we formulated initial domain goals and visual analysis tasks.\\
\textbf{(Phase II) Prototype development and iterative refinement.}
In the subsequent seven months, we focused on the development and refinement of PhenoFlow. We first developed an initial system and presented it to the neurologists (\textbf{E1 and} \textbf{E2}). They were encouraged to explore the system and articulate their thoughts through think-aloud methods \cite{van1994think}. Based on their feedback, we iteratively refined PhenoFlow.

\vspace{-1mm}
\subsection{Current Analysis Workflow and Limitations}
\vspace{-1mm}
Stroke research, mirroring the clinical diagnosis process \cite{elstein1978medical, weber1993determinants}, follows a five-stage process. The first three stages include data wrangling, cohort interpretation, and hypothesis generation. In the subsequent stages, neurologists delve into data exploration and explore insights via descriptive statistics. This cycle of data exploration and insight generation is iteratively repeated until a working hypothesis is established.

In Phase I, we identified that the data wrangling, patient cohort interpretation, and iterative data exploration were significant bottlenecks within the neurologists' analysis workflow. Neurologists explained that, while descriptive statistics offer valuable information, they may lack intuitiveness. Conversely, visualizations provide an intuitive grasp, but interpreting their meaning, particularly in the context of large datasets, presents a considerable challenge. Although previous works have aimed to facilitate interpretation using familiar visual representations \cite{cho2014stroscope, loorak2015timespan}, the growing size and complexity of datasets also escalate the cognitive load associated with interpreting visual encodings. Furthermore, efficient visual encodings for irregularly spaced data (i.e., BP data) are lacking. Neurologists emphasized the need for a more intuitive visual analytics tool that allows for visual exploration of large medical datasets with reduced cognitive load.

\vspace{-1mm}
\subsection{Domain Goals} \label{sec:domain goals}
\vspace{-1mm}
In collaboration with experts, we have identified three domain goals: \\
\textbf{(G1) Define and explore patient cohorts with various patient attributes but with less cognitive load}: Acute ischemic stroke analysis requires the examination of multiple patient attributes to interpret patient outcomes. However, neurologists are not data wranglers or analysts. Therefore, as data volumes and research question complexity increase, this process becomes a significant bottleneck. \\
\textbf{(G2) Identify both recurring and abnormal patterns in irregular BP data}: The main attribute in acute ischemic stroke research is BP data \cite{lawes2004blood, lindenstrom1995influence}. Unlike typical time series data, BP data cannot be continuously and evenly measured. While statistical methods exist to handle such data, visually identifying meaningful patterns in these irregular data remains challenging. \\
\textbf{(G3) Provide clinical evidence with derived outcomes}: Neurologists aim to present clinical evidence alongside the tool's results. For instance, the visualization of a sudden spike in BP might be dismissed as a temporary or stress-related response. However, if such a pattern coincides with specific clinical events, such as recurrence of conditions or medication administration, it could significantly strengthen the insight that these spike patterns indicate a worsening of the patient's condition.

\vspace{-1mm}
\subsection{Visual Analysis Tasks}
\vspace{-1mm}
We translated domain goals into specific visual analysis tasks and validated them in Phase I. The first four tasks (T1-T4) were directly derived from our discussions on domain goals. The last one (T5) was added during Phase II. These five tasks ensure that the design of PhenoFlow is seamlessly aligned with neurologists' needs.\\
\textbf{T1. Iteratively define, filter, and refine patient cohorts.} Constructing patient cohorts is crucial in acute ischemic stroke research. Similar to the information-seeking mantra, neurologists often start their exploration by grouping patients who share similar phenotypes into cohorts. Through this process, they establish, refine, and validate their hypotheses. Acute ischemic stroke is a complex disease influenced by multiple factors that can affect patient outcomes. As a result, the process of constructing and refining cohorts often involves multiple iterations \textbf{(G1)}.\\
\textbf{T2. Compare multiple BP trajectories.}  BP trajectories unveil the health transition of patients or cohorts in acute ischemic stroke. Neurologists often need to examine them in a detailed view and compare them with a comparison group, either between patients or cohorts. The objective of this task is to identify distinctive or similar patterns in BP trajectory and facilitate further exploration.  \textbf{(G2)}.\\
\textbf{T3. Summarize BP trajectories of multiple patients within a cohort.} Neurologists are interested in identifying the health transition pattern, primarily derived from BP trajectories, for patients within a cohort. BP trajectories play a crucial role  in answering clinical questions (e.g., "Has a patient's BP reached a target value?"). Thus, summarizing the BP trajectories of multiple patients in a target cohort is essential for understanding the cohort's characteristics \textbf{(G1)}.\\
\textbf{T4. Display clinical evidence with BP trajectory.}  As mentioned earlier, BP trajectory is a crucial factor in interpreting patients' health status in acute ischemic stroke research. However, it is hard to reach clinical insights solely relying on it. Reaching clinical insights often requires further analysis of multiple clinical factors. Therefore, to reinforce the reliability and interpretability of findings, clinical evidence should be displayed alongside BP trajectories \textbf{(G3)}.\\  
\textbf{T5. Iteratively define and manage sub-cohorts.} Neurologists often engage in multiple iterations of cohort construction and refinement. This process typically begins with the creation of cohorts, which are primarily informed by medical knowledge. Then they partition these cohorts into smaller sub-cohorts based on various clinical attributes, such as outcomes, risk factors, and BP trajectories. Effective management of sub-cohorts is particularly crucial for narrowing the scope of research and refining hypotheses. Therefore, maintaining the context of these sub-cohorts and facilitating their construction is essential.

\vspace{-1mm}
\subsection{Design Requirements} \label{sec:design_requirements}
\vspace{-1mm}
Based on our observations and feedback from experts, we listed key seven design requirements to guide the design of PhenoFlow.\\ 
\textbf{R1. Incorporate familiar visual representations.} Experts acknowledged the necessity of new visualizations to extract insights from irregular, large-scale datasets. However, they also emphasized that these visualizations should not impose additional cognitive loads. As a result, we have incorporate familiar visual representations, such as bars and lines, in the design of PhenoFlow. Despite this, during Phase II, we discovered that neurologists still suffer significant cognitive loads when engaging in visual exploration. This discovery prompted us to define the next design requirement. \\
\textbf{R2. Enable exploration in a more natural manner.} During Phase II, we observed that neurologists experienced cognitive load during the data wrangling process, which includes cohort construction and refinement. For instance, when constructing and reviewing a cohort with multiple conditions, such as \textit{"Male patients over 50 years of age, diagnosed with LAA or SVO, and elevated systolic blood pressure (SBP)>180",} experts had to select multiple conditions and review the results with multiple visualizations. Addressing this process interactively and visually was mentally demanding and sometimes led to confusion. Instead, they expressed a preference for handling the data wrangling process through natural language interactions and focusing on visual exploration of clinical questions. This design requirement led us to develop a human-LLM collaboration workflow. \\
\textbf{R3. Visualize all patient trends in a single view.} Neurologists expressed the need for a comprehensive view of BP patterns within a cohort. Although summarized trends can offer a global overview of the target cohort, irregularities in measurements may introduce bias towards patients with more data points. To address this concern, they requested us to visually represent the data density in this view.\\
\textbf{R4. Integrate descriptive statistics with visualization.} During the iterative exploration process, neurologists typically develop a preliminary understanding of constructed cohorts using raw-level data, such as distribution, variance, and standard deviations. These data enable them to quickly and roughly validate their hypotheses. Therefore, they required a view that combines visualization and descriptive statistics.\\
\textbf{R5. Facilitate easy identification of anomalous data.} Neurologists often focus on values outside the normal range, which may signal anomalous patient statuses. Therefore, clear visual identification of outliers is essential to quickly detect potential issues and make informed decisions regarding patient care.\\
\textbf{R6. Visualize recurring spatio-temporal patterns.} In retrospective studies, it's challenging to ascertain the exact patient situations at specific times. However, during Phase II, \textbf{\textit{E2}} observed that the frequency of BP measurements may reflect the patient's health condition. For instance, a consistent, low-frequency pattern of measurements may suggest the stability of a patient's condition, while an irregular, high-frequency pattern could indicate instability. To enhance the credibility of such findings, it's important to consider not only the magnitude of BP fluctuations (i.e., spatial patterns) but also the timing of measurements (i.e., temporal patterns). This led us to develop a slice-and-wrap visualization, which will be elaborated on in the following sections. \\
\textbf{R7. Enable flexible interactions.} Neurologists often start their exploration by identifying specific ranges of interest both in BP and time. However, these conditions can vary significantly among patients. To address this variability, they require flexible interaction techniques that allow for adjusting the range of interest. Additionally, at the cohort level, patients may exhibit varying clinical attributes or outcomes. Experts have expressed the need to filter out data that shows weak correlations with their research interests.
\vspace{-1.25mm}
\section{Design of PhenoFlow}
\vspace{-1mm}
PhenoFlow is a visual analytics tool designed specifically for acute ischemic stroke patients. It enables a wider range of data exploration by leveraging a Large Language Model (LLM) as a data wrangler. In the previous workflow, neurologists had to fulfill both the roles of data wranglers and researchers. However, in our proposed workflow, neurologists can concentrate on clinical decision-making while the LLM handles the data wrangling tasks.

\vspace{-1.5mm}
\subsection{Human-LLM Collaboration Workflow (R1 and R2)} \label{sec:LLMs}
\vspace{-0.5mm}
While previous research has explored the use of Large Language Models (LLMs) as medical task solvers \cite{hamidi2023evaluation, wang2023can, samaan2023assessing, oh2023chatgpt, holmes2023evaluating, duong2023analysis}, LLMs still exhibit hallucinations in medical domain tasks \cite{thapa2023chatgpt}. Furthermore, the precise performance of LLMs in this domain remains uncertain. Therefore, in designing the human-LLM collaboration workflow, we aimed to keep many possibilities open and collaborated with neurologists to address various medical tasks using the LLM. We will briefly discuss our exploratory phase and then explain the final design of our workflow.

\textbf{Exploratory Phase:} This research began in early 2023 when there was no prior work testing the performance of LLMs in medical tasks. Initially, we collaborated iteratively with neurologists to identify five distinct medical research tasks: (1) generating medical hypotheses from data, (2) recommending literature for medical research, (3) suggesting research methods for validating hypotheses, (4) performing medical data wrangling tasks, and (5) generating and interpreting visualizations. We utilized GPT4-32k for these tasks and employed few-shot prompting to guide the LLMs. To address complex medical knowledge, we initially used retrieval-augmented generation (RAG) that included prior medical research papers. However, due to bias, retrieval errors, and computational overhead, we later discontinued the use of RAG. To reduce hallucination and enable inspection, we adopted a self-reflection approach \cite{shinn2024reflexion}. We utilized the PICO (Population, Intervention, Comparison, Outcomes) hypotheses and research protocols from our collaborators' published studies for testing. Our initial goal was to utilize LLMs as universal companions in medical research.

\textbf{Results:} We found that LLMs struggled with more creative and expansive tasks, such as (1) generating medical hypotheses and (2) recommending literature for medical research. When provided with the PICO hypothesis as input and asked to recommend a medical hypothesis and relevant literature, LLMs exhibited hallucinations in high-temperature settings (e.g., suggesting non-existent literature and generating PICO hypotheses involving incorrect knowledge).
In low-temperature settings, LLMs only modified part of the variables in the PICO hypothesis. However, LLMs demonstrated impressive performance in (3) research method recommendation tasks. When presented with the PICO hypothesis, LLMs successfully recommended research methods similar to those used in actual studies. This result surprised our collaborator (\textbf{E2}), but in terms of reproducibility, LLMs did not consistently yield the same results. These findings posed two significant challenges: \textbf{reproducibility (i.e., consistency)} and \textbf{explainability}.

In the (4) data wrangling tasks, LLMs demonstrated much better results than in previous tasks. We discovered that LLMs are capable of extracting data that meet user requirements from large and intricate datasets. Furthermore, with proper instructions, they can handle missing data by leveraging existing information. However, we encountered another critical challenge in these results: the \textbf{privacy issue}. Lastly, in (5) generating and interpreting visualization tasks, LLMs still struggled to address complex visualizations effectively. Similarly to data wrangling tasks, privacy concerns also exist in these tasks.

\textbf{Final Design:} The prior results led us to design a collaborative workflow incorporating a LLM as a data wrangler. The LLM addresses three medical data wrangling tasks: (1) standardizing inconsistent terminologies, (2) extracting pertinent data from large, complex datasets, and (3) providing interpretable explanations. Ideally, these tasks would be undertaken by a professional data wrangler or with comprehensive documentation of the data. However, such resources are often unavailabl or impractical to employ in real-world clinical settings.

To ensure reproducibility, we employed an \textbf{LLM-based code generation approach} with few-shot prompting to synthesize executable code from user requests. This executable code ensures the consistency of the derived results. To offer explainability for the yielded results, we utilized a combination of natural language explanations and small multiples. The \textbf{natural language explanations} are derived through self-reflection during the reasoning process, while the \textbf{small multiple visualizations} are generated based on the synthesized executable code. We believe these two approaches are complementary, as explanations in natural language are concise but may not be as intuitive, while explanations in visualizations may take up more space but are more easily understood. Lastly, to address privacy issues, we only utilize metadata (e.g., column names, data types, and field coding information) for handling user requests. This approach leverages in-context learning, which refers to the ability of LLMs to learn from and adapt to the context provided in the input. This metadata acts as a form of in-context information that the LLM uses to better understand the user's request and generate more accurate and relevant outputs.

\vspace{-1mm}
\subsubsection{Natural Language Cohort Construction} In PhenoFlow, natural language cohort construction involves both cohort definition and refinement. When experts provide their requests in natural language, the LLM data wrangler goes through a four-step process to conduct data wrangling. In the first stage, the LLM \textbf{normalizes inconsistent terminologies}. These inconsistencies can be caused by users' requests or the data itself. For example, if neurologists want to construct a cohort of patients who have undergone intra-arterial thrombolysis, they can use abbreviations such as IAT, ambiguous terminology (e.g., thrombolysis therapy), or specific procedure codes.

Next, the LLM tries to \textbf{identify the region of interest (ROI)} within data and fields to achieve the user's requirement by utilizing provided metadata as in-context information. Modern medical datasets are often very large and split into multiple files. Therefore, the ROI refers not only to the target field but also to the data location. Then, the LLM starts \textbf{query inference}. Query inference is the final step before generating executable codes. In this stage, the LLM performs multi-step reasoning (i.e., self-reflection) with the data generated in previous stages (e.g., normalized terminologies, metadata, and ROI). The inferences generated in this process are further utilized as natural language explanations in the visual inspection view. Based on the query inference, the LLM \textbf{generates executable code} to extract relevant patient data from multiple data sources and fields. To guide the LLM in synthesizing code, we employed few-shot prompting techniques in this stage. PhenoFlow extracts field names and data from the generated code to create small multiples that will be included in the visual inspection view.

\vspace{-1mm}
\subsubsection{Visual Inspection View} The visual inspection view enables neurologists to examine and understand the result derived by the LLM data wrangler. Neurologists first expand the visual inspection view by clicking the button – for space efficiency, PhenoFlow hides this view by default. After expanding the view, they can inspect 1) their initial natural language request, 2) each code utilized for constructing cohorts, and 3) related inferences and visualizations. Then, neurologists can visually inspect related fields of the cohorts. If they find that the LLM data wrangler performed incorrectly, they can search again by modifying the derived query manually.

\textbf{Design Justification:} Neurologists are familiar with interpreting and validating natural language clinical explanations. Therefore, they can efficiently inspect the LLM data wrangler's behavior with natural language. However, we found that they missed errors when the related condition becomes complex and generates a long explanation. We also tried to compress the length of the natural language explanation with prompt engineering. However, this approach leads to the loss of necessary information and generates hallucinations due to overcompression. Also, there is a risk that these explanations may include hallucinations.

To address this issue, we employed visualization of relevant fields and data. Because these visualizations are generated with the data itself, they do not include hallucinations. Moreover, they are intuitive and easily understandable. To reduce space consumption, we also implemented visualizations that include multiple field information. However, similar to natural language explanation, this visualization also imposes extra cognitive load for neurologists when the condition becomes complex. Therefore, our final design generates visualization per each field and data by utilizing familiar visual encodings (\textbf{R1}). This small multiple approach consumes more space but can provide an intuitive understanding for neurologists.

\vspace{-1.5mm}
\subsubsection{Group View (R4)}
\vspace{-0.75mm}
The group view includes the control panel and the attribute view. The control panel enables neurologists to add the defined cohort to the cohort view, change the BP data type, and manipulate the BP baseline and cycle, which will be utilized in the cohort view. The attribute view allows experts to understand the general characteristics of defined cohorts. To efficiently utilize space and visualize each attribute in one holistic view, we first grouped each attribute into descriptive statistics and the distribution of each BP data type. Each attribute is then displayed in a collapsible view. The distribution of each BP data is visualized with an area chart, while the descriptive statistics are displayed as a table.

\begin{figure*}[tb]
\centering
\includegraphics[width=\textwidth]{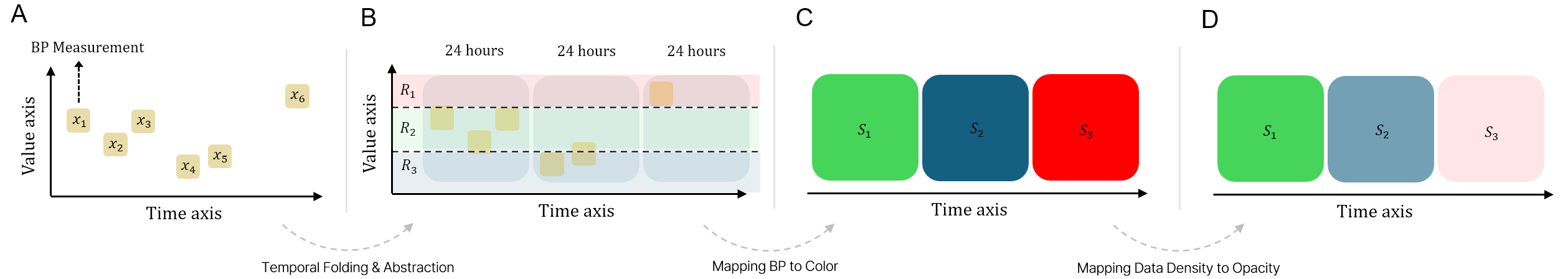}
\vspace{-0.7cm}
\caption{The process of summarizing a patient's BP trajectory. (A) Each patient's BP measurements are irregularly spaced over time. (B) Temporal folding and abstraction are applied to summarize the BP trajectory. (C) The BP range is then mapped to a color channel, and (D) data density is mapped to opacity to reveal the frequency of the measurements.}
\vspace{-2.5mm} 
\end{figure*}

\begin{figure*}[tb]
\centering
\includegraphics[width=\textwidth]{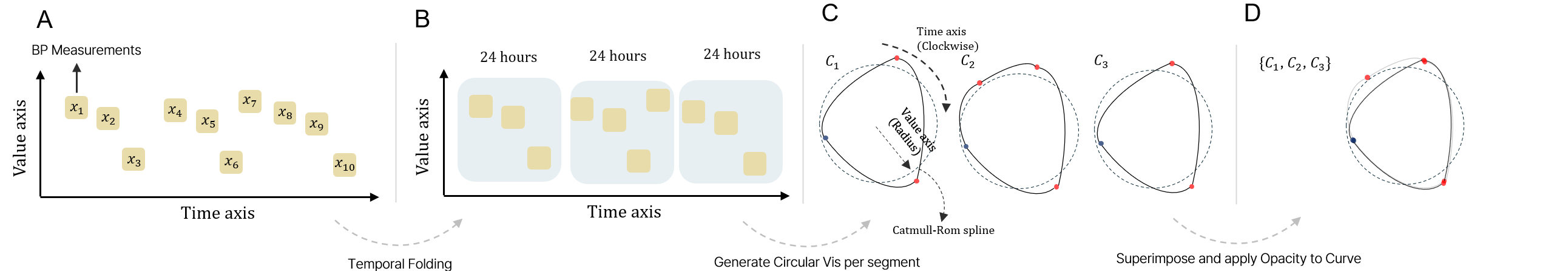}
\vspace{-6mm}
\caption{The process of summarizing a patient's BP trajectory using the slice-and-wrap visualization technique. (A) Each patient's BP measurements are irregularly spaced over time. (B) Temporal folding and abstraction create fixed-size segments (e.g., 24 hours) from the irregular data. (C) A circular visualization is generated for each segment, mapping time to angle and BP value to radius. (D) Data points within each segment are connected using the Centripetal Catmull-Rom spline with opacity applied to the curves. Then, circular visualizations are superimposed to reveal recurring temporal patterns in the patient's BP trajectory.} 
\vspace{-7mm} 
\end{figure*}

\vspace{-1mm}
\subsection{Cohort View} 
\vspace{-0.5mm}
The cohort view enables users to explore and analyze patient data at both the cohort and individual levels. It consists of three main components: (1) the cohort transition view, which summarizes the context of the ongoing analysis and the relationships between cohorts and sub-cohorts; (2) the patient overview, which provides a holistic view of each patient's BP trajectories within a selected cohort; and (3) the slice-and-wrap visualization, which facilitates the identification of distinct and recurring patterns within irregularly spaced BP measurements.

\vspace{-1mm}
\subsubsection{Cohort Transition View (R1 and R2)} 
\vspace{-0.5mm}
The visual exploration process involves the iterative construction and refinement of cohorts. Therefore, the goal of the cohort transition view is to visually summarize the relationships between cohorts and sub-cohorts. To facilitate the iterative refinement of cohorts, we employ natural language cohort construction (i.e., filtering) and use a node-link visualization to summarize the relationships between cohorts. In this view, the initially defined and added cohort becomes the parent node, while the refined sub-cohorts become child nodes. The edges between them represent the relationships. This view aids in understanding the context of the exploration and enables efficient refinement of cohorts.

\vspace{-1mm}
\subsubsection{Patient Overview (R1 and R3)} 
\vspace{-0.5mm}
The patient overview summarizes the BP trajectories of each patient in the selected cohort. To concisely visualize each trajectory in one holistic view, we use a matrix-based visualization. We first apply temporal folding and abstraction to each patient's irregularly spaced BP measurements (Fig. 2B). The applied time window is 24 hours by default, but experts can manipulate the size of the time window with the control panel according to their research goals. We then calculate the mean BP value of each time window and map the BP range to the color channel (Fig. 2C). The range of BP is based on the standards of the American Heart Association but can also be changed with the legend setting in the header menu. Lastly, to visually reveal the data density, we apply opacity to each cell (Fig. 2D). As a result, the irregularly spaced BP measurements can be concisely summarized in the matrix.

The area chart above the matrix represents the entire BP data distribution within a selected cohort. The x-axis represents time in the selected cycle, and the y-axis represents data frequency. By moving the range, experts can identify and filter out the range that has sparse data density. The circle to the left of the UID represents the outcome of each patient. The color is mapped to the discharge mRS by default; however, we also support 3-month mRS and initial NIHSS. Users can change the criteria in the header menu. The bar to the left of the circle represents the numeric clinical variable of each patient (e.g., age, delay, IA surgery time, and mean BP of the nth time window). Experts can sort patients by these criteria and explore similar or distinct patients.

\vspace{-1mm}
\subsubsection{Slice-and-Wrap Visualization (R6)} \label{sec:circular_vis}
\vspace{-0.5mm}
To reveal recurring patterns within irregular data, we propose slice-and-wrap visualization that employs overlaid circular visualization. Similar to the matrix visualization in the patient overview, we apply temporal folding to the patient's BP data (Fig. 3B). However, unlike the matrix visualization, we do not aggregate BP points; instead, we treat these data as time-series data, and construct the circular visualization for each segment (Fig. 3C). The degree of the circle is mapped to the time, and the radius represents the BP value. The color represents whether the BP value is above or below the baseline, and the dotted circle represents the baseline BP. Then, we draw the trend between BP measurements by using the Centripetal Catmull-Rom spline. After that, we overlay these circular visualizations at one point and apply opacity to the curve to visually filter out less frequent patterns while highlighting recurring patterns (Fig. 3D). We utilized superposition because it is useful to reveal similar patterns within the target visualizations \cite{caruso2017creating}.

\textbf{Design Justification:} Applying a similar mechanism to a linear chart is also possible. However, the linear layout does not continuously represent the data, which may hinder the perception of the data's periodic nature. Conversely, the circular layout enables the continuous representation of patterns in a space-efficient way \cite{nusrat2019tasks, chen2021rotate}. One key concern regarding superposition is visual clutter \cite{viola2017pondering}. Similarly, our overlaid circular visualization may have a similar issue. However, BP data is typically measured less than 5 times a day, except in special cases, so the data density is not high. Moreover, we only visualize the BP trajectory in this visualization to reduce the visual clutter. The rest of the clinical data will be displayed in a linear bar chart.

\vspace{-2mm}
\subsubsection{Linear Bar Chart (R5 and R7)}
\vspace{-1mm}
At the left of the slice-and-wrap visualization, we juxtaposed a linear layout bar chart. The linear layout bar chart visualizes the selected patient's BP measurements and clinical information (e.g., IV, IA surgery time, and recurrence). The baseline of the bar chart represents the target BP value, which typically indicates a normal BP value (e.g., 120 mmHG for SBP). The color of each bar represents whether the patient's BP value is above or below the baseline. Since patients' BP varies individually, we designed the visualization to allow moving the baseline to facilitate visual exploration. If the user right-clicks, they can change the baseline to the dual range mode. The dual range mode provides interactive upper and lower baselines. The dual baseline filters out BP values that do not fall into the target range. In summary, the linear bar chart visualizes the BP trajectory with familiar visual representations and enables users to find distinct BP patterns (e.g., sudden spikes and sharp drops) within irregularly spaced BP measurements.

\vspace{-2.5mm}
\section{Evaluation}
\vspace{-1mm}
In this section, we present the results of case studies conducted with four experienced neurologists. 
{Prior to the hour-long case study session, we introduced PhenoFlow to participants by using a demo scenario. The demo scenario only utilized patients’ demographic information. Then, participants were encouraged to explore the CRCS-K dataset \cite{bae2022david} while articulating their thoughts through think-aloud methods \cite{van1994think}. Further feedback was gathered through post-study interviews.} None of the experts were collaborators or co-authors of this paper.

\vspace{-2mm}
\subsection{Case Study I: Evaluating the Human-LLM Collaboration Workflow}
\vspace{-2mm}
This case study assessed the reliability and efficiency of the human-LLM collaboration workflow in facilitating the analysis by experts with no prior knowledge of the dataset. For an objective assessment of performance, we did not provide a codebook or any information detailing the meaning of each data field. The study was conducted separately with \textbf{P1} and \textbf{P2}, neurologists with nine years of experience who had never analyzed the CRCS-K dataset before. \textbf{P1}'s research interests involve investigating health outcomes of elderly patients who have experienced stroke due to large artery atherosclerosis (LAA) (\textbf{Cohort \#1 [C1]}). Similarly, \textbf{P2}'s research interests include investigating the health outcomes of patients who have experienced a stroke due to LAA and the administration of medications. As both experts share similar research interests and the results of their analyses were comparable, this section will primarily focus on the findings from \textbf{P1}'s study.

\label{sec:evaluation}

\begin{figure}[tb]
\centering
\includegraphics[width=\columnwidth]{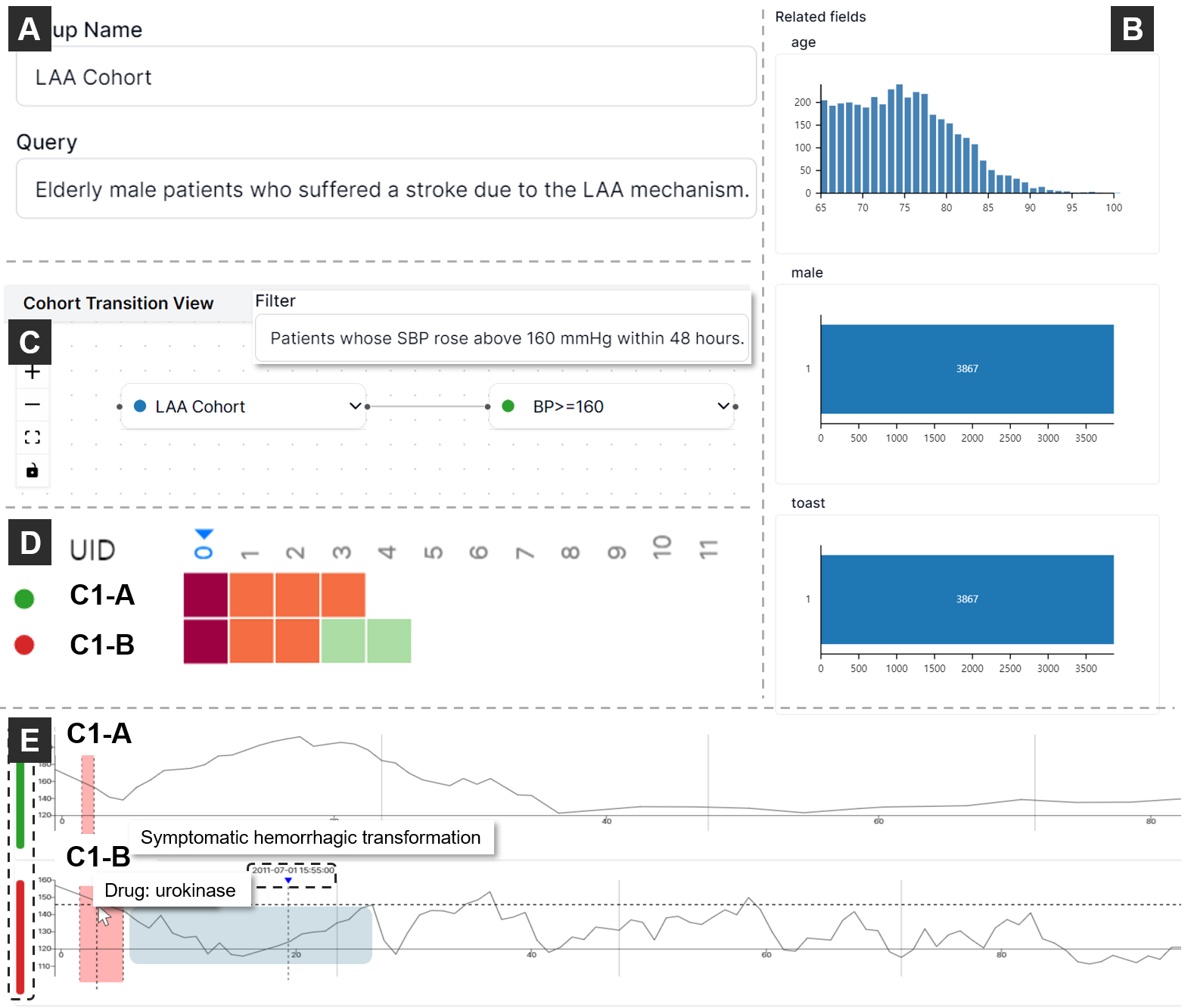}
\vspace{-6mm} 
\caption{Case Study I - (A) \textbf{P1} began her exploration by defining a cohort using a natural language query. (B) Through small multiples in the Inspection View, she verified that the results obtained by the LLM data wrangler aligned with her requirements. (C) By iteratively refining the cohort with natural language filters, she identified (D) two patients of interest in the matrix. (E) By comparing the two patients' data in detail and utilizing the tooltip (i.e., urokinase) and event marker (i.e., Sym HT), she discovered a potential influencing factor for their disparate outcomes.}
\vspace{-7.5mm} 
\end{figure}

\textbf{Defining cohorts with natural language.} Without prior knowledge of the dataset, \textbf{P1} began by submitting her requests in natural language (Fig. 4A). Her query was "\textit{Elderly male patients who suffered a stroke due to the LAA.}" This query, which included abbreviations (e.g., LAA) and potentially ambiguous terms (e.g., elderly), was suitable for evaluating PhenoFlow's ability to handle complex, real-world queries.

\textbf{Understanding the defined cohorts.} PhenoFlow quickly identified 3,867 patients matching the first query (\textbf{C1}). Given her prior experience with GPT-4 encountering hallucinations in medical research, \textbf{P1} was curious whether the resulting cohorts accurately reflected her requests. She utilized the visual inspection view for further examination (Fig. 4B). PhenoFlow's query inferences for \textbf{C1} stated: "\textit{The user's request specifies 'elderly male patients', typically considered to be 65 years or older. Consequently, the data query filters for male patients (male==1) aged 65 and above (age>=65) who have had a stroke with LAA etiology (toast==1).}" The Visual Debugging View provided visualizations of patient distributions across fields, impressing \textbf{P1} with PhenoFlow's ability to construct cohorts from natural language queries and to provide understandable explanations. 
{With traditional methods, she typically spent anywhere from tens of minutes to an hour creating and validating a cohort. However, with PhenoFlow, she was able to create and validate the cohort in just a few minutes, significantly reducing the time and effort required.} She confirmed that these query conditions aligned with her research needs, and proceeded with further analysis.

\textbf{Iteratively refining cohorts.} Upon reviewing the cohorts, \textbf{P1} noted, "\textit{In cases of acute ischemic stroke, patients generally exhibit elevated BP within 48 hours. The absence of such phenomenon typically indicates a relatively benign condition, necessitating its exclusion from my study.}" Subsequently, she refined \textbf{C1} with an additional natural language instruction: "\textit{Patients whose SBP rose above 160 mmHg within 48 hours.}" She then sorted patients in the refined cohort by BP value on the first day and found two patients in \textbf{C1} who had similar admission conditions but significantly different outcomes.

\textbf{Inspecting patients with different outcomes.} \textbf{P1} realized that both patients (\textbf{C1-A} and \textbf{C1-B}) were admitted within half an hour of stroke onset and had a total admission duration of 4 days. Although there was a slight difference in the initial National Institutes of Health Stroke Scale (NIHSS), both patients were similar in most conditions. However, their outcomes were significantly different: \textbf{C1-A}'s discharge-modified Rankin scale (mRS) was 1, indicating minimal disability, whereas \textbf{C1-B}'s was 5, indicating severe disability.

\textbf{P1} compared both patients' BP trajectories with linear bar chart and found that \textbf{C1-B}'s total surgery time was five times longer than \textbf{C1-A}'s. \textbf{C1-B} had also exhibited a BP drop after IA surgery, while \textbf{C1-A}'s BP had increased after surgery and then steadily decreased after one day. \textbf{P1} stated, "\textit{In the treatment of acute ischemic stroke, medication may be administered to lower the BP immediately, or the BP may be intentionally increased to aid the patient's recovery. Both approaches are applicable in the clinical practice for acute ischemic stroke patients.}" She confirmed that both patients received appropriate treatments. However, she found that within 24 hours after IA surgery, \textbf{C1-B} showed a U-shaped BP pattern with a symptomatic hemorrhagic transformation event, which is bleeding into the brain tissue that worsens symptoms. \textbf{P1} commented, "\textit{When BP shows such U-shaped pattern with symptomatic hemorrhagic transformation, it is a strong sign that the patient's condition is deteriorating.}"

\textbf{Generating data-driven hypothesis and validating it.} She further investigated other clinical factors to find the potential cause of the different outcomes. She discovered that \textbf{C1-B} was given urokinase, a thrombolytic agent to dissolve blood clots, while \textbf{C1-A} only had mechanical thrombectomy without a thrombolytic agent. She stated, "\textit{While urokinase was previously utilized for thrombolytic therapy, its application in intra-arterial (IA) surgery raises the risk of bleeding. Recently, device-assisted procedures have become a preferred approach.}" She hypothesized that the use of a thrombolytic agent might be a cause of the different outcomes. Taking all the evidence together, she concluded, "\textit{\textbf{C1-B} received urokinase during IA surgery and experienced a drop in BP, followed by subsequent BP elevation and a symptomatic hemorrhagic transformation. Administration of urokinase during IA surgery is not used much these days due to concerns about hemorrhagic side effects. This may explain the divergent outcomes of \textbf{C1-A} and \textbf{C1-B}, despite their similar initial conditions.}"

Similar to \textbf{P1}'s results, \textbf{P2} also favored natural language cohort construction and found multiple patients who revealed U-shaped BP patterns with recurrence events. He also confirmed that urokinase is a strong medication and has the adverse effect of vascular weakening. Therefore, the U-shaped BP pattern accompanied by symptomatic hemorrhagic transformation events can be seen as a negative health transition pattern, and it is highly likely that the administration of such medication is the cause of this pattern.

\vspace{-3mm}
\subsection{Case Study II - Validating the Slice-And-Wrap Design and Discovering Triangular BP Patterns}
\vspace{-1.5mm}
In this study, we aimed to validate that the design of PhenoFlow aligns with clinical knowledge and that the visually explored findings connect to clinical evidence. We conducted the study with \textbf{P3}, who had 13 years of experience, and \textbf{P4}, who had 12 years of experience. Both experts were professors of neurology at university hospitals. \textbf{P3} had prior research experience with the CRCS-K dataset, while \textbf{P4} did not. \textbf{P3} commented, "\textit{Compared to other datasets, the CRCS-K dataset is large, and the fields are recorded in detail. However, when I conducted prior research with this dataset, I took a considerable amount of time to understand the codebook due to the overload of information.}"

\vspace{-0.5mm}
\textbf{Constructing a cohort using natural language and inspection.} Introduced to PhenoFlow's human-LLM collaboration workflow, they were intrigued and decided to explore new research ideas by visually exploring this large dataset. After discussing potential natural language queries, \textbf{P4} suggested, "\textit{Since antiplatelet agents are usually given as standard, let's construct a cohort of patients who showed SBP >= 180 mmHg and received an antiplatelet agent within 48 hours.}" PhenoFlow processed the user's natural language query and constructed a target cohort. However, after inspecting the results with the visual inspection view, they found that the LLM data wrangler requires an additional data field - 'antiplatelet therapy administration' to address the user's request. \textbf{P3}, who had prior analysis experience with the CRCS-K dataset, recalled, "\textit{Ah, the CRCS-K dataset does not have separate records for the timing of antiplatelet agent administration, which might explain these results.}" After identifying the cause of the error, they revised their query to "\textit{Patients who showed SBP >= 180 mmHg within 48 hours.}" PhenoFlow went through the reasoning process again and constructed a cohort (\textbf{Cohort \#2 [C2]}) that met the users' requirements.

\textbf{Visually exploring the constructed cohort.} They proceeded to visually explore the constructed cohort by utilizing the cohort overview. They first sorted patients by mean BP on day 0. At the top of the matrix, they found a patient who showed a sustained high BP pattern (i.e., SBP >= 180 mmHg) for 8 days since initial hospitalization (\textbf{C2-A}). However, the discharge mRS score of \textbf{C2-A} was 1 (i.e., the green color), which indicated minimal disability. \textbf{P4} commented, "\textit{This patient showed a very high BP sequence over 8 days. However, unless the patient has a specific medical condition, the preferred approach these days is to wait for the BP to drop naturally rather than aggressively lowering it. Moreover, this patient was in his 40s, which is considered young.}" They hypothesized that \textbf{C2-A}'s good outcome despite his long-standing high BP was likely due to his young age. Aligned with their hypothesis, they found that \textbf{C2-A}'s BP sequence gradually stabilized from day 9 of hospitalization and returned to the normal range 3 days before discharge. However, they felt that a more cautious exploration was needed before drawing conclusions, and they decided to explore the BP changes in detail through the slice-and-wrap visualization.

\begin{figure}[tb]
\centering
\includegraphics[width=\columnwidth]{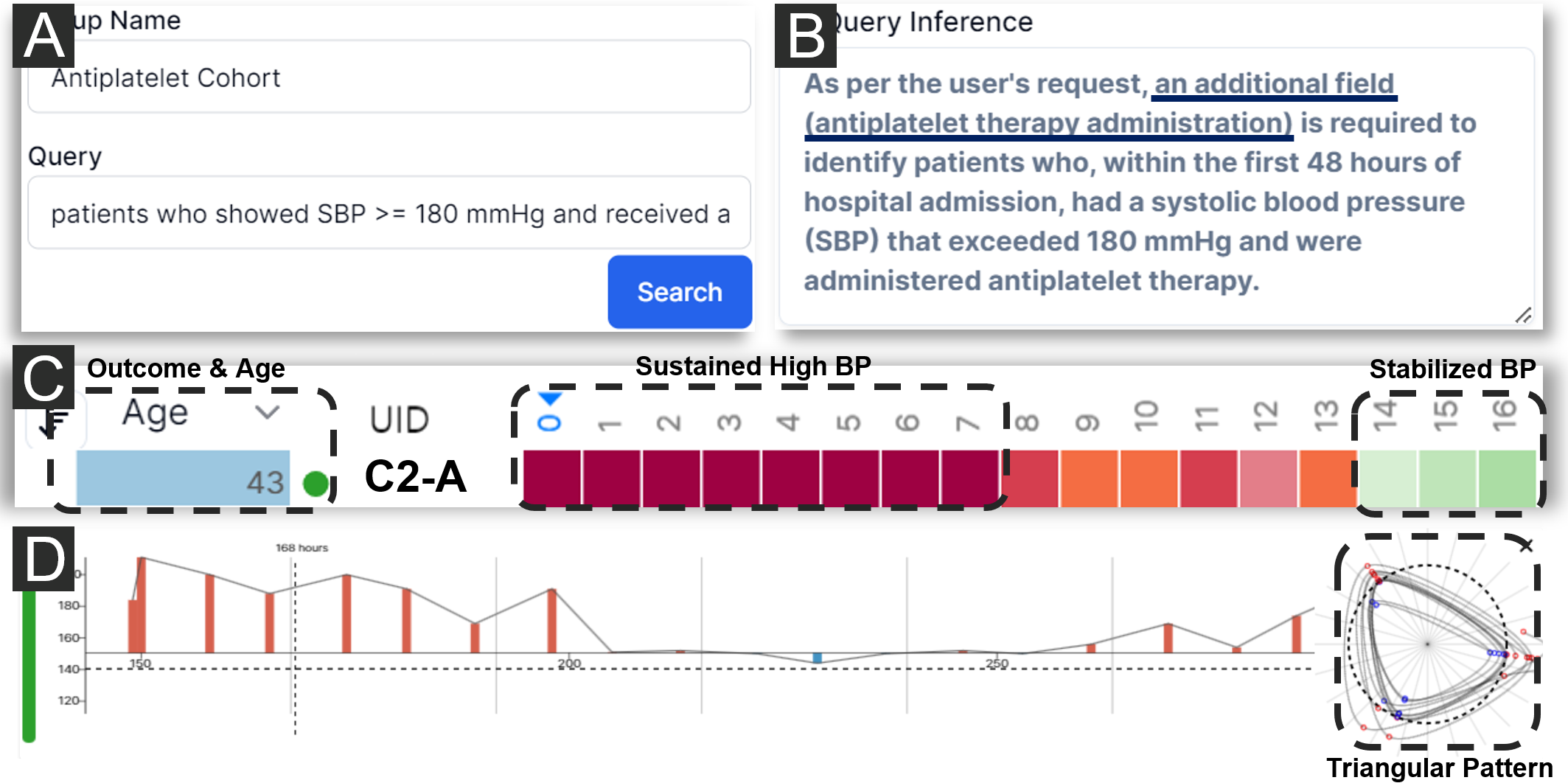}
\vspace{-6.5mm}
\caption{Case Study II - (A) After defining the target cohort, (B) experts found that the LLM data wrangler needed an additional field to meet the user's request. (C) After modifying their query, they identified a patient with sustained high BP for 8 days. (D) Examining the patient's BP trajectory using the slice-and-wrap visualization, they discovered a triangular pattern that potentially indicated the patient's stabilized condition.}
\vspace{-7.5mm}
\end{figure}

\textbf{Exploring the BP trajectory in detail.} They first inspected \textbf{C2-A}'s BP trajectory with a linear bar chart. Because \textbf{C2-A} maintained a high BP range for most of his admission period, they increased the baseline to 150 mmHg. Similar to the BP sequence in the patient overview, they found that \textbf{C2-A}'s BP reached the target BP (150 mmHg) on day 9. However, when they inspected \textbf{C2-A}'s BP with the slice-and-wrap visualization, they 
{visually} found a triangular-shaped BP pattern starting from day 6 
{at a glance}. \textbf{P3} stated, \textit{"Most hospitals measure BP at regular times of the day - morning, noon, and evening - once a patient's condition has stabilized. So this triangular pattern can also be seen as an indication that the patient's condition has stabilized."} However, \textbf{P4} suspected that this regular pattern might not be due to the patient's stable condition, but rather to the hospital's protocols. Therefore, they established a new hypothesis – the triangular pattern reveals the patient's stabilized status – and further compared the BP trajectories of patients from different healthcare organizations.

\textbf{Comparing BP patterns between different organizations.} They selected 6 patients from different organizations and compared their BP trajectories. Contrary to their thought, this triangular pattern appeared regardless of the organization, although there were slight variations in shape. \textbf{P3} commented, \textit{"While there are slight variations in the shape, given the timing of these patterns, this triangular pattern may reflect the patient's stabilized state."} After further discussing the implications of these patterns, they concluded that the slice-and-wrap design visually reflects both the meaning of BP variance and clinical information.
\vspace{-2mm}
\section{Future Directions} \label{sec:future_direction}
\vspace{-1mm}
The series of case studies demonstrates that PhenoFlow empowers neurologists to derive meaningful insights and expand their exploration range in the complex and extensive medical dataset. Despite the promising results, we found there are still limitations and areas for future research. We further discuss future directions for integrating LLM into visual analytics tasks within the medical domain.

\vspace{-2mm}
\subsection{LLM as a Medical Data Wrangler}
\vspace{-1mm}
Most visual analytics systems assume that the target data has undergone some degree of data wrangling, but unfortunately, the majority of real-world medical datasets do not align with this assumption. Numerous experts have stated that data wrangling for medical data is messy, repetitive, and resource-intensive, requiring a significant amount of manpower. Furthermore, the current data wrangling methods for medical datasets struggle to keep up with the rapid accumulation of data. This challenge persists even when neurologists analyze the wrangled data. Given that they are not data analysts, extracting relevant insights from these extensive and complex datasets to answer their research questions becomes a daunting task. To tackle this challenge, we employed LLM as a data wrangler, with the LLM data wrangling capability emerging as one of the most appreciated features of PhenoFlow. 
{Although we have not formally studied the impact of varying data quality on performance, through iterative testing of the LLM data wrangler with the extensive CRCS-K dataset, we believe this workflow can be extended to other diseases or domains with large structured datasets.}

However, we want to emphasize that LLM is a ``fragile'' data wrangler, possessing inherent power but facing challenges in 1) directing its unpredictable behavior, 2) understanding and explaining its actions, and 3) ensuring consistent outcomes. Prompt engineering emerges as the most effective means to influence LLM, but a universally reliable 'magic' prompt for medical tasks remains elusive. To ensure verifiability and credibility, LLM workflows should be designed with multiple reasoning steps while keeping the design concise. While granting LLMs more freedom (i.e., high temperature), constructing long task chains, or incorporating complex modules may enhance their performance on complex medical tasks, it also raises the risk of unpredictable behavior. 
{Additionally, the highly disorganized metadata or data structure may pose difficulties for the LLMs during the inference step.} Therefore, we advocate for integrating LLMs in workflows where 1) results can be validated, 2) tasks don't demand a high level of complexity or creativity, and 3) LLMs can either reduce repetitive tasks for domain experts or augment their expertise like data wrangling.

\vspace{-2mm}
\subsection{Rethinking the Use of Visualizations in Medical Data Analysis}
\vspace{-1.5mm}
Domain experts have emphasized the growing importance of visualization in clinical research papers. However, through our collaborations and case studies, we found that the concept of visualization remains unfamiliar to many experts in this field. Specifically, due to their limited ability to interpret and understand visualizations, the full potential of visualization in this domain remains untapped. Traditionally, medical visual analytics has relied on familiar visual encodings and low-complexity visualizations to minimize the cognitive load on domain experts. Yet, with the increasing volume and complexity of data, this conventional approach is no longer sufficient.

Moreover, we argue that the current visual analytic workflow overly relies on domain experts' limited ability to interpret visualizations, which in turn fosters their distrust in the visual output. Indeed, we have observed numerous instances where domain experts, upon discovering interesting findings through visualization, initially question the credibility of these findings and subsequently seek to proceed directly to the statistical analysis phase. As the interpretation of visualization relies on the experts' perception, they struggle to place trust in visually derived results. Therefore, we advocate for research within the visualization community to develop technologies that enhance domain experts' visual interpretation capabilities. We believe that these advancements, coupled with effective visualizations, are pivotal in harnessing the full potential of visualization within the medical domain.

As discussed in section 7.1, LLMs can complement the limited capabilities of domain experts by simplifying visualizations and supplementing their expertise. PhenoFlow demonstrates how LLMs can reduce the cognitive load  of visual analytics by narrowing down the exploration range of extensive datasets. However, we believe that LLMs can extend their utility further by either generating visualizations or interpreting complex visualizations through interactive dialogues with experts. Essentially, LLMs can serve as artificial data analysts, augmenting domain experts' ability to interpret and leverage visualizations. Exploring this direction will be a promising avenue for future research.
\vspace{-5mm}
\vspace{-1mm}
\section{Conclusion}
\vspace{-1mm}
In this paper, we identified three key domain goals for analyzing data for acute ischemic stroke patients through a year-long collaboration with expert neurologists. We additionally explored opportunities to integrate LLMs into the visual analytic workflow. Based on the findings, we develop PhenoFlow—a visual analytics system driven by an innovative human-LLM collaborative workflow. Utilizing LLMs as data wranglers, PhenoFlow reduces cognitive load on domain experts, allowing them to concentrate on clinical decision-making. Furthermore, PhenoFlow introduces the slice-and-wrap visualization technique, which facilitates the exploration of meaningful patterns in irregularly spaced BP data. 
{This technique could potentially be extended to other chronic diseases that exhibit irregularities and require long-term monitoring, such as diabetes and heart diseases.}
To validate our tool's efficacy, we conducted case studies involving four experienced neurologists. 
{The limited number of participants precluded a quantitative evaluation of cognitive load reduction. However, the qualitative} results demonstrate PhenoFlow's effectiveness in supporting iterative analysis and unveiling important insights from large, complex real-world stroke datasets. While acknowledging its limitations and challenges, PhenoFlow represents a significant advancement in the realm of visual analytics tools incorporating LLMs for navigating extensive and intricate medical datasets. With further research and refinement, we envision this approach significantly enhancing clinicians' interaction with medical data and deriving meaningful insights, ultimately contributing to better patient outcomes.
\section*{Supplemental Materials}
All supplemental materials are available on OSF at \href{https://osf.io/q6yc4/}{https://osf.io/q6yc4/}. The repository includes: (1) a full paper, (2) the source code of PhenoFlow, including prompts, and (3) additional figures and a video demonstrating the core features of PhenoFlow.
\section*{Acknowledgements}
This work was partly supported by the National Research Foundation of Korea (NRF) grant funded by the Korea government (MSIT) (No. RS-2023-00251406), the Hankuk University of Foreign Studies Research Fund, and the Institute of Information \& communications Technology Planning \& Evaluation (IITP) grant funded by the Korea government (MSIT) [No. RS-2021-II211343, Artificial Intelligence Graduate School Program (Seoul National University)]. The authors extend their gratitude to the researchers at CRCS-K (Clinical Research Collaboration for Stroke in Korea) for their invaluable support and for providing access to the data collected by the CRCS-K stroke registry, which significantly contributed to this study. Additionally, the authors thank Seokhyeon Park for his valuable feedback that improved the user interface of PhenoFlow.

\bibliographystyle{abbrv-doi-hyperref}

\bibliography{reference}

\begin{thebibliography}{10}

\bibitem{achiam2023gpt}
J.~Achiam, S.~Adler, S.~Agarwal, L.~Ahmad, I.~Akkaya, F.~L. Aleman, D.~Almeida, J.~Altenschmidt, S.~Altman, S.~Anadkat, et~al.
\newblock Gpt-4 technical report.
\newblock {\em arXiv preprint arXiv:2303.08774}, 2023.

\bibitem{adams1993classification}
H.~P. Adams~Jr, B.~H. Bendixen, L.~J. Kappelle, J.~Biller, B.~B. Love, D.~L. Gordon, and E.~Marsh~3rd.
\newblock Classification of subtype of acute ischemic stroke. definitions for use in a multicenter clinical trial. toast. trial of org 10172 in acute stroke treatment.
\newblock {\em stroke}, 24(1):35--41, 1993.

\bibitem{aigner2007visualizing}
W.~Aigner, S.~Miksch, W.~M{\"u}ller, H.~Schumann, and C.~Tominski.
\newblock Visualizing time-oriented data—a systematic view.
\newblock {\em Computers \& Graphics}, 31(3):401--409, 2007.

\bibitem{albers2002transient}
G.~W. Albers, L.~R. Caplan, J.~D. Easton, P.~B. Fayad, J.~Mohr, J.~L. Saver, and D.~G. Sherman.
\newblock Transient ischemic attack—proposal for a new definition, 2002.

\bibitem{bae2022david}
H.-J. Bae.
\newblock David g. sherman lecture award: 15-year experience of the nationwide multicenter stroke registry in korea.
\newblock {\em Stroke}, 53(9):2976--2987, 2022.

\bibitem{batal2009temporal}
I.~Batal, L.~Sacchi, R.~Bellazzi, and M.~Hauskrecht.
\newblock A temporal abstraction framework for classifying clinical temporal data.
\newblock In {\em AMIA Annual Symposium Proceedings}, vol. 2009, p.~29. American Medical Informatics Association, 2009.

\bibitem{brown2020language}
T.~Brown, B.~Mann, N.~Ryder, M.~Subbiah, J.~D. Kaplan, P.~Dhariwal, A.~Neelakantan, P.~Shyam, G.~Sastry, A.~Askell, et~al.
\newblock Language models are few-shot learners.
\newblock {\em Advances in neural information processing systems}, 33:1877--1901, 2020.

\bibitem{caban2015visual}
J.~J. Caban and D.~Gotz.
\newblock Visual analytics in healthcare--opportunities and research challenges.
\newblock {\em Journal of the American Medical Informatics Association}, 22(2):260--262, 2015.

\bibitem{carlini2022quantifying}
N.~Carlini, D.~Ippolito, M.~Jagielski, K.~Lee, F.~Tramer, and C.~Zhang.
\newblock Quantifying memorization across neural language models.
\newblock {\em arXiv preprint arXiv:2202.07646}, 2022.

\bibitem{carlini2021extracting}
N.~Carlini, F.~Tramer, E.~Wallace, M.~Jagielski, A.~Herbert-Voss, K.~Lee, A.~Roberts, T.~Brown, D.~Song, U.~Erlingsson, et~al.
\newblock Extracting training data from large language models.
\newblock In {\em 30th USENIX Security Symposium (USENIX Security 21)}, pp. 2633--2650, 2021.

\bibitem{caruso2017creating}
V.~Caruso, A.~Cattaneo, and J.-L. Gurtner.
\newblock Creating technology-enhanced scenarios to promote observation skills of fashion-design students.
\newblock {\em Form@ re-Open Journal per la formazione in rete}, 17(1):4--17, 2017.

\bibitem{chen2021rotate}
K.-T. Chen, T.~Dwyer, B.~Bach, and K.~Marriott.
\newblock Rotate or wrap? interactive visualisations of cyclical data on cylindrical or toroidal topologies.
\newblock {\em IEEE Transactions on Visualization and Computer Graphics}, 28(1):727--736, 2021.

\bibitem{cho2014stroscope}
M.~Cho, B.~Kim, H.-J. Bae, and J.~Seo.
\newblock Stroscope: Multi-scale visualization of irregularly measured time-series data.
\newblock {\em IEEE transactions on visualization and computer graphics}, 20(5):808--821, 2014.

\bibitem{dobbins2023leafai}
N.~J. Dobbins, B.~Han, W.~Zhou, K.~F. Lan, H.~N. Kim, R.~Harrington, {\"O}.~Uzuner, and M.~Yetisgen.
\newblock Leafai: query generator for clinical cohort discovery rivaling a human programmer.
\newblock {\em Journal of the American Medical Informatics Association}, 30(12):1954--1964, 2023.

\bibitem{du2016coping}
F.~Du, B.~Shneiderman, C.~Plaisant, S.~Malik, and A.~Perer.
\newblock Coping with volume and variety in temporal event sequences: Strategies for sharpening analytic focus.
\newblock {\em IEEE transactions on visualization and computer graphics}, 23(6):1636--1649, 2016.

\bibitem{duong2023analysis}
D.~Duong and B.~D. Solomon.
\newblock Analysis of large-language model versus human performance for genetics questions.
\newblock {\em European Journal of Human Genetics}, pp. 1--3, 2023.

\bibitem{elstein1978medical}
A.~S. Elstein, L.~S. Shulman, and S.~A. Sprafka.
\newblock {\em Medical problem solving: An analysis of clinical reasoning}.
\newblock Harvard University Press, 1978.

\bibitem{federico2014qualizon}
P.~Federico, S.~Hoffmann, A.~Rind, W.~Aigner, and S.~Miksch.
\newblock Qualizon graphs: Space-efficient time-series visualization with qualitative abstractions.
\newblock In {\em Proceedings of the 2014 international working conference on advanced visual interfaces}, pp. 273--280, 2014.

\bibitem{guo2022medical}
Q.~Guo, S.~Cao, and Z.~Yi.
\newblock A medical question answering system using large language models and knowledge graphs.
\newblock {\em International Journal of Intelligent Systems}, 37(11):8548--8564, 2022.

\bibitem{guo2021survey}
Y.~Guo, S.~Guo, Z.~Jin, S.~Kaul, D.~Gotz, and N.~Cao.
\newblock Survey on visual analysis of event sequence data.
\newblock {\em IEEE Transactions on Visualization and Computer Graphics}, 28(12):5091--5112, 2021.

\bibitem{hamidi2023evaluation}
A.~Hamidi and K.~Roberts.
\newblock Evaluation of ai chatbots for patient-specific ehr questions.
\newblock {\em arXiv preprint arXiv:2306.02549}, 2023.

\bibitem{he2023survey}
K.~He, R.~Mao, Q.~Lin, Y.~Ruan, X.~Lan, M.~Feng, and E.~Cambria.
\newblock A survey of large language models for healthcare: from data, technology, and applications to accountability and ethics.
\newblock {\em arXiv preprint arXiv:2310.05694}, 2023.

\bibitem{holmes2023evaluating}
J.~Holmes, Z.~Liu, L.~Zhang, Y.~Ding, T.~T. Sio, L.~A. McGee, J.~B. Ashman, X.~Li, T.~Liu, J.~Shen, et~al.
\newblock Evaluating large language models on a highly-specialized topic, radiation oncology physics.
\newblock {\em Frontiers in Oncology}, 13, 2023.

\bibitem{jauch2013guidelines}
E.~C. Jauch, J.~L. Saver, H.~P. Adams~Jr, A.~Bruno, J.~Connors, B.~M. Demaerschalk, P.~Khatri, P.~W. McMullan~Jr, A.~I. Qureshi, K.~Rosenfield, et~al.
\newblock Guidelines for the early management of patients with acute ischemic stroke: a guideline for healthcare professionals from the american heart association/american stroke association.
\newblock {\em Stroke}, 44(3):870--947, 2013.

\bibitem{johnson2020mimic}
A.~Johnson, L.~Bulgarelli, T.~Pollard, S.~Horng, L.~A. Celi, and R.~Mark.
\newblock Mimic-iv.
\newblock {\em PhysioNet. Available online at: https://physionet. org/content/mimiciv/1.0/(accessed August 23, 2021)}, pp. 49--55, 2020.

\bibitem{johnson2016mimic}
A.~E. Johnson, T.~J. Pollard, L.~Shen, L.-w.~H. Lehman, M.~Feng, M.~Ghassemi, B.~Moody, P.~Szolovits, L.~Anthony~Celi, and R.~G. Mark.
\newblock Mimic-iii, a freely accessible critical care database.
\newblock {\em Scientific data}, 3(1):1--9, 2016.

\bibitem{lammarsch2013interactive}
T.~Lammarsch, W.~Aigner, A.~Bertone, M.~B{\"o}gl, T.~Gschwandtner, S.~Miksch, and A.~Rind.
\newblock Interactive visual transformation for symbolic representation of time-oriented data.
\newblock In {\em Human-Computer Interaction and Knowledge Discovery in Complex, Unstructured, Big Data: Third International Workshop, HCI-KDD 2013, Held at SouthCHI 2013, Maribor, Slovenia, July 1-3, 2013. Proceedings}, pp. 400--419. Springer, 2013.

\bibitem{lawes2004blood}
C.~M. Lawes, D.~A. Bennett, V.~L. Feigin, and A.~Rodgers.
\newblock Blood pressure and stroke: an overview of published reviews.
\newblock {\em Stroke}, 35(3):776--785, 2004.

\bibitem{lindenstrom1995influence}
E.~Lindenstr{\o}m, G.~Boysen, and J.~Nyboe.
\newblock Influence of systolic and diastolic blood pressure on stroke risk: a prospective observational study.
\newblock {\em American journal of epidemiology}, 142(12):1279--1290, 1995.

\bibitem{loorak2015timespan}
M.~H. Loorak, C.~Perin, N.~Kamal, M.~Hill, and S.~Carpendale.
\newblock Timespan: Using visualization to explore temporal multi-dimensional data of stroke patients.
\newblock {\em IEEE transactions on visualization and computer graphics}, 22(1):409--418, 2015.

\bibitem{lyi2020comparative}
S.~LYi, J.~Jo, and J.~Seo.
\newblock Comparative layouts revisited: Design space, guidelines, and future directions.
\newblock {\em IEEE Transactions on Visualization and Computer Graphics}, 27(2):1525--1535, 2020.

\bibitem{mistry2019blood}
E.~A. Mistry, H.~Sucharew, A.~M. Mistry, T.~Mehta, N.~Arora, A.~K. Starosciak, F.~De~Los Rios La~Rosa, J.~E. Siegler~III, N.~R. Barnhill, K.~Patel, et~al.
\newblock Blood pressure after endovascular therapy for ischemic stroke (best) a multicenter prospective cohort study.
\newblock {\em Stroke}, 50(12):3449--3455, 2019.

\bibitem{moskovitch2009medical}
R.~Moskovitch and Y.~Shahar.
\newblock Medical temporal-knowledge discovery via temporal abstraction.
\newblock In {\em AMIA annual symposium proceedings}, vol. 2009, p. 452. American Medical Informatics Association, 2009.

\bibitem{murdoch2013inevitable}
T.~B. Murdoch and A.~S. Detsky.
\newblock The inevitable application of big data to health care.
\newblock {\em Jama}, 309(13):1351--1352, 2013.

\bibitem{narayan2022can}
A.~Narayan, I.~Chami, L.~Orr, S.~Arora, and C.~R{\'e}.
\newblock Can foundation models wrangle your data?
\newblock {\em arXiv preprint arXiv:2205.09911}, 2022.

\bibitem{nusrat2019tasks}
S.~Nusrat, T.~Harbig, and N.~Gehlenborg.
\newblock Tasks, techniques, and tools for genomic data visualization.
\newblock In {\em Computer Graphics Forum}, vol.~38, pp. 781--805. Wiley Online Library, 2019.

\bibitem{oh2023chatgpt}
N.~Oh, G.-S. Choi, and W.~Y. Lee.
\newblock Chatgpt goes to the operating room: evaluating gpt-4 performance and its potential in surgical education and training in the era of large language models.
\newblock {\em Annals of Surgical Treatment and Research}, 104(5):269, 2023.

\bibitem{plaisant2003lifelines}
C.~Plaisant, R.~Mushlin, A.~Snyder, J.~Li, D.~Heller, and B.~Shneiderman.
\newblock Lifelines: using visualization to enhance navigation and analysis of patient records.
\newblock In {\em The craft of information visualization}, pp. 308--312. Elsevier, 2003.

\bibitem{qiu2023large}
J.~Qiu, L.~Li, J.~Sun, J.~Peng, P.~Shi, R.~Zhang, Y.~Dong, K.~Lam, F.~P.-W. Lo, B.~Xiao, et~al.
\newblock Large ai models in health informatics: Applications, challenges, and the future.
\newblock {\em IEEE Journal of Biomedical and Health Informatics}, 2023.

\bibitem{rastegar2018biocreative}
M.~Rastegar-Mojarad, S.~Liu, Y.~Wang, N.~Afzal, L.~Wang, F.~Shen, S.~Fu, and H.~Liu.
\newblock Biocreative/ohnlp challenge 2018.
\newblock In {\em Proceedings of the 2018 ACM International Conference on Bioinformatics, Computational Biology, and Health Informatics}, pp. 575--575, 2018.

\bibitem{rind2011visually}
A.~Rind, W.~Aigner, S.~Miksch, S.~Wiltner, M.~Pohl, F.~Drexler, B.~Neubauer, and N.~Suchy.
\newblock Visually exploring multivariate trends in patient cohorts using animated scatter plots.
\newblock In {\em Ergonomics and Health Aspects of Work with Computers: International Conference, EHAWC 2011, Held as Part of HCI International 2011, Orlando, FL, USA, July 9-14, 2011. Proceedings}, pp. 139--148. Springer, 2011.

\bibitem{samaan2023assessing}
J.~S. Samaan, Y.~H. Yeo, N.~Rajeev, L.~Hawley, S.~Abel, W.~H. Ng, N.~Srinivasan, J.~Park, M.~Burch, R.~Watson, et~al.
\newblock Assessing the accuracy of responses by the language model chatgpt to questions regarding bariatric surgery.
\newblock {\em Obesity surgery}, 33(6):1790--1796, 2023.

\bibitem{scheer2022visualization}
J.~Scheer, A.~Volkert, N.~Brich, L.~Weinert, N.~Santhanam, M.~Krone, T.~Ganslandt, M.~Boeker, and T.~Nagel.
\newblock Visualization techniques of time-oriented data for the comparison of single patients with multiple patients or cohorts: Scoping review.
\newblock {\em Journal of medical Internet research}, 24(10):e38041, 2022.

\bibitem{sedlmair2012design}
M.~Sedlmair, M.~Meyer, and T.~Munzner.
\newblock Design study methodology: Reflections from the trenches and the stacks.
\newblock {\em IEEE transactions on visualization and computer graphics}, 18(12):2431--2440, 2012.

\bibitem{shi2024ehragent}
W.~Shi, R.~Xu, Y.~Zhuang, Y.~Yu, J.~Zhang, H.~Wu, Y.~Zhu, J.~Ho, C.~Yang, and M.~D. Wang.
\newblock Ehragent: Code empowers large language models for complex tabular reasoning on electronic health records.
\newblock {\em arXiv preprint arXiv:2401.07128}, 2024.

\bibitem{shinn2024reflexion}
N.~Shinn, F.~Cassano, A.~Gopinath, K.~Narasimhan, and S.~Yao.
\newblock Reflexion: Language agents with verbal reinforcement learning.
\newblock {\em Advances in Neural Information Processing Systems}, 36, 2024.

\bibitem{singhal2023large}
K.~Singhal, S.~Azizi, T.~Tu, S.~S. Mahdavi, J.~Wei, H.~W. Chung, N.~Scales, A.~Tanwani, H.~Cole-Lewis, S.~Pfohl, et~al.
\newblock Large language models encode clinical knowledge.
\newblock {\em Nature}, 620(7972):172--180, 2023.

\bibitem{singhal2023towards}
K.~Singhal, T.~Tu, J.~Gottweis, R.~Sayres, E.~Wulczyn, L.~Hou, K.~Clark, S.~Pfohl, H.~Cole-Lewis, D.~Neal, et~al.
\newblock Towards expert-level medical question answering with large language models.
\newblock {\em arXiv preprint arXiv:2305.09617}, 2023.

\bibitem{soni2023quehry}
S.~Soni, S.~Datta, and K.~Roberts.
\newblock quehry: a question answering system to query electronic health records.
\newblock {\em Journal of the American Medical Informatics Association}, 30(6):1091--1102, 2023.

\bibitem{thapa2023chatgpt}
S.~Thapa and S.~Adhikari.
\newblock Chatgpt, bard, and large language models for biomedical research: opportunities and pitfalls.
\newblock {\em Annals of biomedical engineering}, 51(12):2647--2651, 2023.

\bibitem{trummer2022codexdb}
I.~Trummer.
\newblock Codexdb: Synthesizing code for query processing from natural language instructions using gpt-3 codex.
\newblock {\em Proceedings of the VLDB Endowment}, 15(11):2921--2928, 2022.

\bibitem{van1994think}
M.~Van~Someren, Y.~F. Barnard, and J.~Sandberg.
\newblock The think aloud method: a practical approach to modelling cognitive.
\newblock {\em London: AcademicPress}, 11(6), 1994.

\bibitem{viola2017pondering}
I.~Viola and T.~Isenberg.
\newblock Pondering the concept of abstraction in (illustrative) visualization.
\newblock {\em IEEE transactions on visualization and computer graphics}, 24(9):2573--2588, 2017.

\bibitem{wang2009temporal}
T.~D. Wang, C.~Plaisant, B.~Shneiderman, N.~Spring, D.~Roseman, G.~Marchand, V.~Mukherjee, and M.~Smith.
\newblock Temporal summaries: Supporting temporal categorical searching, aggregation and comparison.
\newblock {\em IEEE transactions on visualization and computer graphics}, 15(6):1049--1056, 2009.

\bibitem{wang2023can}
Z.~Wang, R.~Li, B.~Dong, J.~Wang, X.~Li, N.~Liu, C.~Mao, W.~Zhang, L.~Dong, J.~Gao, et~al.
\newblock Can llms like gpt-4 outperform traditional ai tools in dementia diagnosis? maybe, but not today.
\newblock {\em arXiv preprint arXiv:2306.01499}, 2023.

\bibitem{weber1993determinants}
E.~U. Weber, U.~B{\"o}ckenholt, D.~J. Hilton, and B.~Wallace.
\newblock Determinants of diagnostic hypothesis generation: effects of information, base rates, and experience.
\newblock {\em Journal of Experimental Psychology: Learning, Memory, and Cognition}, 19(5):1151, 1993.

\bibitem{zhang2018idmvis}
Y.~Zhang, K.~Chanana, and C.~Dunne.
\newblock Idmvis: Temporal event sequence visualization for type 1 diabetes treatment decision support.
\newblock {\em IEEE transactions on visualization and computer graphics}, 25(1):512--522, 2018.

\end{thebibliography}

\end{document}